\documentclass[fleqn,usenatbib]{mnras}

\usepackage{newtxtext,newtxmath}

\usepackage[T1]{fontenc}

\DeclareRobustCommand{\VAN}[3]{#2}
\let\VANthebibliography\thebibliography
\def\thebibliography{\DeclareRobustCommand{\VAN}[3]{##3}\VANthebibliography}

\usepackage{graphicx}	
\usepackage{amsmath}	
\usepackage[dvipsnames]{xcolor}
\usepackage[capitalise, nameinlink]{cleveref}

\crefname{section}{§}{§§}
\Crefname{section}{§}{§§}

\title[Flattening of the lag spectrum with increasing luminosity]{Kilohertz quasi-periodic oscillations in neutron-star X-ray binaries: Flattening of the lag spectrum with increasing luminosity}

\author[V. Peirano et al.]{
Valentina Peirano,$^{1}$\thanks{E-mail: v.peirano@astro.rug.nl}
Mariano M\'endez$^{1}$
\\
$^{1}$Kapteyn Astronomical Institute, University of Groningen, P.O. BOX 800, 9700 AV Groningen, The Netherlands\\
}

\date{Accepted XXX. Received YYY; in original form ZZZ}

\pubyear{2020}

\begin{document}
\label{firstpage}
\pagerange{\pageref{firstpage}--\pageref{lastpage}}
\maketitle

\begin{abstract}
We study the energy-dependent time lags and rms fractional amplitude of the kilohertz quasi-periodic oscillations (kHz QPOs) of a group of neutron-star low mass X-ray binaries (LMXBs). We find that for the lower kHz QPO the slope of the best-fitting linear model to the time-lag spectrum and the total rms amplitude integrated over the 2 to 25 keV energy band both decrease exponentially with the luminosity of the source. For the upper kHz QPO the slope of the time-lag spectrum is consistent with zero, while the total rms amplitude decreases exponentially with the luminosity of the source. We show that both the slope of the time-lag spectrum and the total rms amplitude of the lower kHz QPO are linearly correlated with a slope of $\sim 1$. Finally, we discuss the mechanism that could be responsible for the radiative properties of the kHz QPOs, with the variability originating in a Comptonising cloud or corona that is coupled to the innermost regions of the accretion disc, close to the neutron star.
\end{abstract}




\section{Introduction}
Quasi-periodic oscillations (QPOs) are variability features observed in the power spectra of neutron-star and black-hole Low-Mass X-ray Binaries \citep[LMXBs;][]{strohmayerMillisecondXRayVariability1996,vanderklisDiscoverySubmillisecondQuasiperiodic1996, morganRXTEObservationsQPOs1997,strohmayerDiscovery450HZ2001}. The oscillations detected at the highest frequencies in the power density spectra of LMXBs are called kilohertz QPOs \citep[kHz QPOs; see][for a review]{mendezHighfrequencyVariabilityNeutronstar2020}. In neutron-star LMXBs, kHz QPOs are sometimes observed in pairs and, depending on their central frequency, are called the lower and the upper kHz QPOs. Besides these high-frequency features, the power spectra of LMXBs have a plethora of different types of variability and a noise component, known as broad-band noise (BBN), at lower frequencies \citep[see e.g.,][for a description of variability features in the power spectra of LMXBs]{belloniUnifiedDescriptionTiming2002}. 
 
Variability in LMXBs is usually fitted using a sum of Lorentzian functions that describe the shape of the power spectrum of a source \citep[e.g.][]{miyamotoXrayVariabilityGX1991,groveTimingNoiseProperties1994,bergerComparisonFastTiming1998,nowakAreThereThree2000,vanstraatenMultiLorentzianTimingStudy2002,pottschmidtLongTermVariability2003}. These Lorentzian functions are characterised by three paramenters: the central frequency, $\nu_0$, the quality factor, $Q = \nu_0/$FWHM, where FWHM is the full-width at half-maximum, and the integral of the power under the Lorentzian function. These properties allow us to identify kHz QPOs, with $\nu_0$ ranging from 250 to 1300 Hz, and differentiate the lower from the upper component. The quality factor \citep[][]{barretAbruptDropCoherence2005, peilleSpectraltimingPropertiesUpper2015} and the rms amplitude  \citep[][]{disalvoStudyTemporalBehavior2001,mendezAmplitudeKilohertzQuasiperiodic2001}, for example, behave differently for the lower and the upper kHz QPO. In particular, for the lower kHz QPO the rms amplitude first increases and then decreases with increasing QPO frequency \citep[see e.g.,][]{vanstraatenMultiLorentzianTimingStudy2002,barretCoherenceKilohertzQuasiperiodic2006,mendezMaximumAmplitudeCoherence2006,ribeiroAmplitudeKilohertzQuasiperiodic2019}, whereas the rms amplitude of the upper kHz QPO generally decreases with increasing QPO frequency \citep[][]{vanstraatenMultiLorentzianTimingStudy2002,mendezMaximumAmplitudeCoherence2006,altamiranoDiscoveryKilohertzQuasiPeriodic2008,troyerSystematicSpectralTimingAnalysis2018}.

Another property of the QPO signal is the energy-dependent time (or phase) lag. In recent years, time lags have been used more often to study the properties of the physical components in the accretion flow that produce the variability in LMXBs \citep[e.g.][]{vaughanDiscoveryMicrosecondTime1997,kaaretDiscoveryMicrosecondSoft1999,barretSOFTLAGSNEUTRON2013,deavellarTimeLagsKilohertz2013,deavellarPhaseLagsQuasiperiodic2016}. Particularly for the lower kHz QPO, the time lags are generally soft (the soft/low-energy photons lag behind the hard/high-energy ones), ranging from $\sim$15 to $\sim$40 $\mu$s between photons in the 3 to 8 keV band and those in the 8 to 30 keV band \citep[][]{deavellarTimeLagsKilohertz2013,barretSOFTLAGSNEUTRON2013}. The time lags of the upper kHz QPO are hard (the hard/high-energy photons lag behind the soft/low-energy ones) for all energy bands and QPO frequencies, inconsistent with those of the lower kHz QPO \citep[][]{deavellarTimeLagsKilohertz2013,peilleSpectraltimingPropertiesUpper2015}.

\begin{table*}
    \centering
    \begin{tabular}{cccccccc}
    \hline
     & \multicolumn{4}{c}{lower kHz QPO} & \multicolumn{3}{c}{upper kHz QPO}\\
    Source & $m$ ($\mu$s/keV) & $c$ ($\mu$s) & $\alpha_1$ (\%/keV) & rms$_{2-25\mathrm{keV}}$ (\%)& $m$ ($\mu$s/keV) & $c$ ($\mu$s) & rms$_{2-25\mathrm{keV}}$ (\%)\\
    \hline
    4U 0614$+$09 & $8.1 \pm 1.0$ & $54.4 \pm 4.9$ & $1.17 \pm 0.03$ & $10.34 \pm 0.25$ & $0.5 \pm 1.1$ & $2.3 \pm 5.5$ & $10.2 \pm 0.4$\\
    4U 1702$-$43 & $6.4 \pm 0.7$ & $52.9 \pm 3.2$ & $0.97 \pm 0.02$ & $8.57 \pm 0.16$ & $5.7 \pm 2.3$ & $44.8 \pm 8.7$ & $8.83 \pm 1.05$\\
    Aql X$-$1 & $5.6 \pm 0.7$ & $42.5 \pm 3.3$ & $0.90 \pm 0.02$ & $7.91 \pm 0.17$ & & & \\
    4U 1608$-$52 & $4.9 \pm 0.6$ & $37.3 \pm 2.6$ & $1.01 \pm 0.02$ & $8.94 \pm 0.14$ & $2.3 \pm 1.2$ & $15.0 \pm 5.4$ & $10.3 \pm 0.7$\\
    4U 1915$-$05 & $4.0 \pm 3.7$ & $35.4 \pm 16.7$ & $1.0 \pm 0.2$ & $8.76 \pm 1.34$ & & & \\
    4U 1636$-$53 & $3.2 \pm 0.3$ & $24.3 \pm 1.3$ & $0.91 \pm 0.01$ & $8.03 \pm 0.11$ & $-0.63 \pm 0.57$ & $-5.0 \pm 2.0$ & $9.6 \pm 0.3$\\
    4U 1728$-$34 & $2.2 \pm 0.4$ & $20.3 \pm 1.4$ & $0.80 \pm 0.01$ & $6.99 \pm 0.12$ & $-0.8 \pm 0.3$ & $-7.2 \pm 1.3$ & $8.3 \pm 0.2$\\
    EXO 1745$-$248 & $1.5 \pm 1.1$ & $11.6 \pm 4.1$ & $0.90 \pm 0.05$ & $7.83 \pm 0.37$ & & & \\
    4U 1820$-$30 & $1.1 \pm 0.4$ & $8.9 \pm 1.7$ & $0.54 \pm 0.02$ & $4.74 \pm 0.12$ & $1.1 \pm 1.7$ & $10.1 \pm 7.0$ & $4.7 \pm 0.4$\\
    4U 1735$-$44 & $0.9 \pm 0.5$ & $8.1 \pm 1.8$ & $0.73 \pm 0.02$ & $6.42 \pm 0.15$ & & & \\
    XTE J1739$-$285 & $0.5 \pm 1.5$ & $4.0 \pm 5.9$ & $0.95 \pm 0.08$ & $8.37 \pm 0.58$ & & & \\
    SAX J1748.9$-$202 & $0.4 \pm 3.6$ & $-20.4 \pm 12.4$ & $1.2 \pm 0.2$ & $10.39 \pm 1.26$ & & & \\
    IGR J17191$-$2821 & $0.08 \pm 1.41$ & $2.7 \pm 5.2$ & $0.91 \pm 0.07$ & $8.00 \pm 0.47$ & & & \\
    4U 1705$-$44 & $-6.5 \pm 3.9$ & $-54.5 \pm 17.9$ & $0.55 \pm 0.08$ & $4.86 \pm 0.58$ & & & \\
    \hline
    \end{tabular}
\caption{Slope, $m$, and intercept, $c$, of the best-fitting linear function to the time-lag energy spectra of the lower and upper kHz QPOs as given by \citet{troyerSystematicSpectralTimingAnalysis2018}. Slope, $\alpha_1$, below the break of the best-fitting broken-line function to the fractional rms energy spectra of the lower kHz QPO in Fig. 4 in \citet{troyerSystematicSpectralTimingAnalysis2018}. Total fractional rms amplitude between 2 and 25 keV, rms$_{2-25\mathrm{keV}}$, of both the lower and upper kHz QPOs. For the broken-line fit to the fractional rms energy spectra of the lower kHz QPO, we fixed $\alpha_2$, the slope above the break, to zero and linked the break energy to be the same for all sources, with a best-fitting value $E_{\mathrm{break}} = 10.3 \pm 0.2$ keV. The slope $m$ is shown here with changed signs with respect to the values given by \citet{troyerSystematicSpectralTimingAnalysis2018} as we considered the linear model fitted to the data to be $\Delta t(E) = -mE + c$.}
    \label{tab:bestfit-lag/rms_vs_energy}
\end{table*}
Correlations derived from the properties of QPOs \citep[see][for a review on timing techniques]{uttleyXRayReverberationAccreting2014} have been subject of detailed study, as the QPO properties are believed to be closely related to phenomena occurring close to the neutron star or the black hole, at the innermost regions of LMXBs \citep[see e.g.,][]{millerEffectsRapidStellar1998, vanderklisQPOPhenomenon2005, psaltisProbesTestsStrongField2008}. Consequently, QPOs properties are a unique tool to directly observe and measure effects that occur in extreme gravitational environments, as the ones in LMXBs. While the central frequency of kHz QPOs reveals the dynamical timescales of the system \citep[][]{strohmayerMillisecondXRayVariability1996}, the lags can be used to understand the radiative processes of the accretion flow at even shorter timescales, closer to the compact object, revealing the radiative mechanism responsible for the variability in LMXB \citep[][]{leeComptonizationQPOOrigins1998}.

Multiple models have been developed throughout the years to try and explain the relations among the different properties of QPOs, and between these properties and the spectral evolution of the source or its mass accretion state. Most models focused their efforts in explaining the dynamical properties of the QPOs, exploring the mechanism that produces the QPO central frequencies \citep{stellaLenseThirringPrecessionQuasiperiodic1997,millerSonicPointModelKilohertz1998,osherovichKilohertzQuasiperiodicOscillations1999,kluzniakPhysicsKHzQPOs2001,titarchukRayleighTaylorGravityWaves2003}. Models that explain the dependence of the $Q$ factor on the central frequency of the QPO have been used less often to describe the nature of the variability \citep[][]{barretAbruptDropCoherence2005}. Finally, models have been proposed to explain the dependence of the lags and rms amplitude on frequency and energy \citep[][]{leeComptonizationQPOOrigins1998,leeComptonUpscatteringModel2001,kumarEnergyDependentTime2014,kumarConstrainingSizeComptonizing2016,karpouzasComptonizingMediumNeutron2020}.

Recently \citet{troyerSystematicSpectralTimingAnalysis2018} carried out a systematic study of the time lags, fractional rms amplitude, covariance and coherence function of the lower and upper kHz QPOs in 14 different LMXBs. These authors observed that the time lags of the lower kHz QPOs were consistently soft on all sources and either decreased or remained more or less constant as the energy increases from $\sim3$ keV up to 12 keV. On the contrary, for the upper kHz QPO the time lags were consistent with zero, with larger errors than for the lags of the lower kHz QPO. \citet{troyerSystematicSpectralTimingAnalysis2018} concluded that this must mean that the mechanisms producing the lower and upper kHz QPOs are different in nature \citep[see also][]{deavellarTimeLagsKilohertz2013,peilleSpectraltimingPropertiesUpper2015}. \citet{mendezMaximumAmplitudeCoherence2006} studied the behaviour of the maximum quality factor and fractional rms amplitude of the kHz QPOs in 12 different LMXBs, and found that for both the lower and upper kHz QPOs the maximum fractional rms amplitude decreases exponentially with increasing luminosity of the source, while for the lower kHz QPOs the maximum coherence increases up to a certain luminosity and then decreases exponentially.
\begin{table}
    \centering
    \begin{tabular}{ccc}
    \hline
    Source & $ l = L/L_{\mathrm{Edd}}$ & Reference\\
    \hline
    4U 0614$+$09 & $0.065 \pm 0.016$ & 1\\
    4U 1702$-$43 & $0.046 \pm 0.018$ & 2\\
    Aql X$-$1 & $0.015 \pm 0.041$ & 1\\
    4U 1608$-$52 & $0.030 \pm 0.075$ & 1\\
    4U 1636$-$53 & $0.085 \pm 0.021$ & 1\\
    4U 1728$-$34 & $0.070 \pm 0.017$ & 1\\
    4U 1820$-$30 & $0.20 \pm 0.05$ & 1\\
    4U 1735$-$44 & $0.12 \pm 0.03$ & 1\\
    \hline
    \end{tabular}
    \caption{Luminosities in Eddington units for the 8 sources in our sub-sample selected from \citet{troyerSystematicSpectralTimingAnalysis2018}. References: (1) \citet{fordSimultaneousMeasurementsXRay2000}; (2) \citet{jonkerDiscoveryKilohertzQuasiperiodic2001}.}
    \label{tab:lum_persource}
\end{table}

Motivated by these results, in this paper we explore the dependence of both the fractional rms amplitude and the time lags of the lower kHz QPOs in the LMXBs studied by \citet{troyerSystematicSpectralTimingAnalysis2018} upon luminosity. In \cref{sec:data} we describe the data that we obtained from the literature. In \cref{sec:results} we show the results of the analysis of these data and in \cref{sec:discussion} we discuss the implications of these results.

\section{Data Selection and Analysis}
\label{sec:data}
The data we use in this paper were collected with the Rossi X-ray Timing Explorer \citep[\textit{RXTE;}][]{bradtXRayTimingExplorer1993}, using the Proportional Counting Array \citep[\textit{PCA;}][]{jahodaInorbitPerformanceCalibration1996}, and taken from the analysis in \citet{troyerSystematicSpectralTimingAnalysis2018}. Using the full \textit{RXTE/PCA} energy band, with a time resolution of at least 125 $\mu$s, they analysed observations of 14 neutron-star LMXBs that showed either one or two kHz QPOs. \citet{troyerSystematicSpectralTimingAnalysis2018} identified the lower and upper kHz QPO in each observation using the criteria of \citet{peilleSpectraltimingPropertiesUpper2015}, based on the quality factor $Q$ of the QPOs. They calculated the energy-dependent spectral timing products using the shift-and-add technique \citep{mendezDiscoverySecondKilohertz1998} on both the power spectra and cross spectra of all observations. Energy-dependent time lags were calculated in several narrow energy bands with respect to the full-reference band, and averaged over frequency across the FWHM of the Lorentzian function that describes the QPO \citep[see Fig. 8 and Fig. 9 in][]{troyerSystematicSpectralTimingAnalysis2018}. Broadband lags were calculated as the average of the frequency-dependent time lags between two broad energy bands, $3-8$ keV and $8-20$ keV \citep[see Fig. 6 and Fig. 7 in][]{troyerSystematicSpectralTimingAnalysis2018}. In the convention of \citet{troyerSystematicSpectralTimingAnalysis2018}, positive time lags represent ‘soft lags', where the low-energy photons arrive after the high-energy photons to the detector \citep[see][for more details about the analysis of the data]{troyerSystematicSpectralTimingAnalysis2018}.

When selecting the sources in our sample, we consider that both fractional rms amplitude \citep[][]{disalvoStudyTemporalBehavior2001,mendezAmplitudeKilohertzQuasiperiodic2001,barretAbruptDropCoherence2005,mendezMaximumAmplitudeCoherence2006,ribeiroRelationPropertiesKilohertz2017,ribeiroAmplitudeKilohertzQuasiperiodic2019} and lags \citep[][]{deavellarTimeLagsKilohertz2013,barretSOFTLAGSNEUTRON2013,peilleSpectraltimingPropertiesUpper2015,deavellarPhaseLagsQuasiperiodic2016} of the lower kHz QPO depend on QPO frequency. \citet{barretSOFTLAGSNEUTRON2013} and \citet{deavellarTimeLagsKilohertz2013} showed, for example, that in both 4U 1608$-$52 and 4U 1636$-$53 the lag of the lower kHz QPO increases and then decreases with increasing QPO frequency. In Fig. 1 in \citet{troyerSystematicSpectralTimingAnalysis2018} it is apparent that from the 14 sources they analysed, only 8 have measurements covering the full range of frequencies of the lower kHz QPOs. As a consequence of the behaviour of the lag and the fractional rms amplitude with QPO frequency, averaging those quantities over all frequencies in the 6 sources from \citet{troyerSystematicSpectralTimingAnalysis2018} that have sparse measurements in the frequency range of the QPO would lead to systematic errors in the analysis. Therefore, here we only study the 8 sources (see \cref{tab:lum_persource}) with data covering evenly the entire frequency range of the lower kHz QPO. In contrast, for the upper kHz QPO, the coverage in QPO frequency is not as critical, since the lags are constant with QPO frequency \citep[see e.g.,][]{deavellarTimeLagsKilohertz2013,peilleSpectraltimingPropertiesUpper2015}. Hence, we also included in our sample the 6 sources where \citet{troyerSystematicSpectralTimingAnalysis2018} identified the presence of upper kHz QPOs.

Once we defined our sample, we searched in the literature for the average luminosity of each source. For most sources we used the lower kHz QPO frequency of each source to read off the luminosity from Fig. 1 in \citet{fordSimultaneousMeasurementsXRay2000}, following \citet{mendezMaximumAmplitudeCoherence2006}. For 4U 1702$-$43 we obtained the luminosity from Figure 4 in \citet{jonkerDiscoveryKilohertzQuasiperiodic2001}, using the fractional rms amplitude as input. In \citet{fordSimultaneousMeasurementsXRay2000} they estimated the luminosity using the 2-50 keV flux of the source as a measure of the bolometric flux and they normalised this quantity by a value of $L_{\mathrm{Edd}}=2.5\times10^{38}$ erg s$^{-1}$, which corresponds to the Eddington luminosity of a 1.9 M$_\odot$ neutron star accreting matter with cosmic abundance \citep[the same method was used in][]{jonkerDiscoveryKilohertzQuasiperiodic2001}. The distances used to calculate the luminosity and the corresponding references are listed in Table 1 in \citet{fordSimultaneousMeasurementsXRay2000}. More modern estimates of the distance to our sources \citep[see e.g.,][]{kuulkersPhotosphericRadiusExpansion2003, gallowayThermonuclearTypeXRay2008,kuulkersWhatIgnitesNeutron2010,arnasonDistancesGalacticXray2021b} are in agreement within errors with the distances used by \citet{fordSimultaneousMeasurementsXRay2000}. For the error of the luminosity we considered two factors: First, for each source, the distribution of the intensity when the lower kHz QPO is present has a spread of 10$-$30\% of the average value \citep[e.g.,][]{mendezAmplitudeKilohertzQuasiperiodic2001,barretSupportingEvidenceSignature2007,barretDiscoveryUpperKilohertz2008}. Second, the error of the luminosity reflects the error in the estimate of the distance to these sources. While distances given in the literature have an accuracy of $\sim$50\%, the values obtained using different methods have a spread of $\sim$20\%, much smaller than the errors of the individual measurements \citep[for a detailed discussion of these and other potential sources of error of the luminosity, see][]{mendezMaximumAmplitudeCoherence2006}. Given the above, here we use a fixed error of 25\% for $L/L_{\mathrm{Edd}}$. The luminosity of each source in our sample and its error is listed in \cref{tab:lum_persource}.
\begin{figure}
    \centering
    \includegraphics[scale=.45]{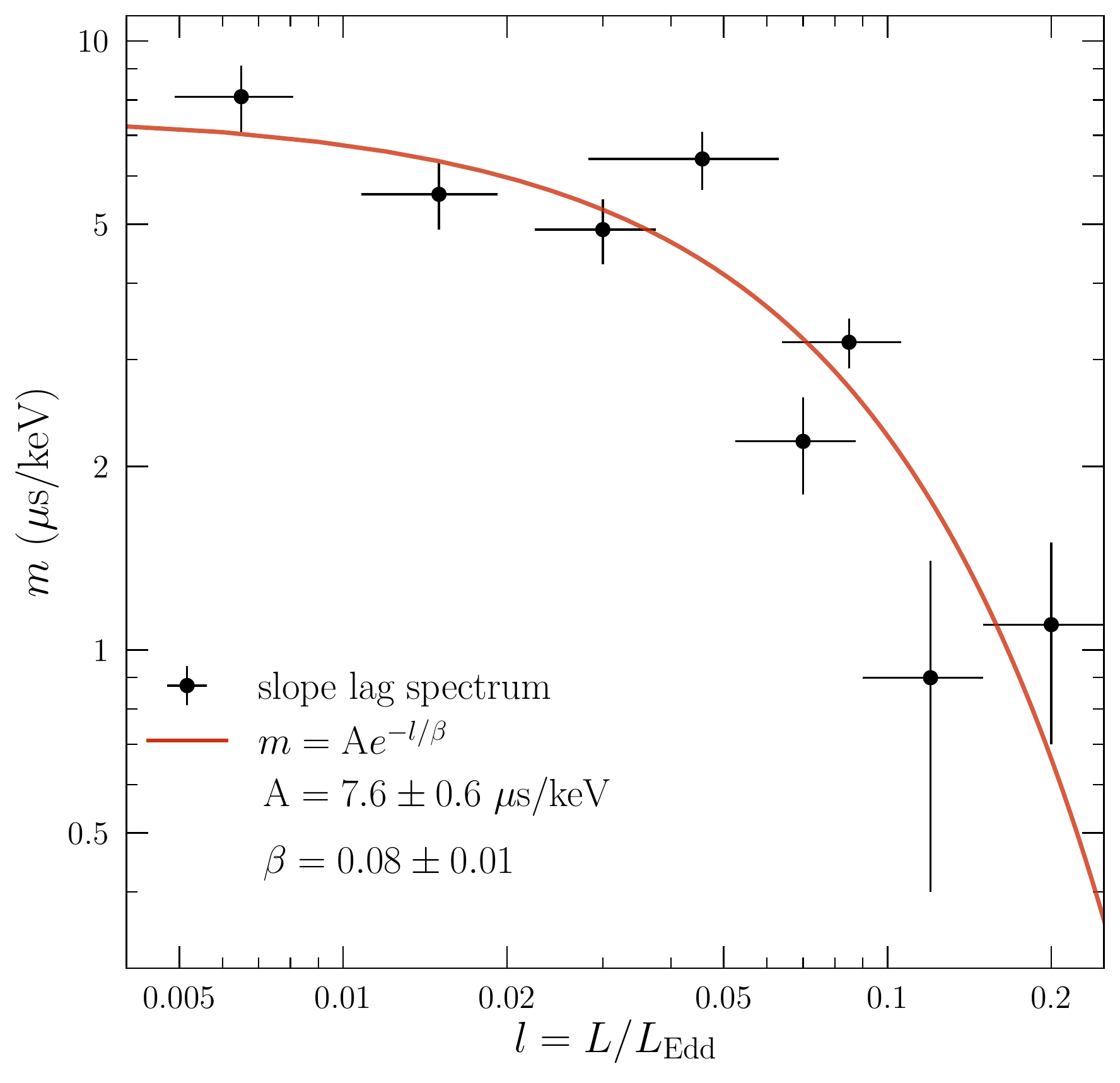}
    \caption{Slope of the time-lag spectrum, $m$ \citep[taken from][]{troyerSystematicSpectralTimingAnalysis2018}, of the lower kHz QPO vs. luminosity in Eddington units for the 8 sources in \cref{tab:lum_persource}. The red solid line indicates the best-fitting exponential model to the data (see text).}
    \label{fig:slopelag_luminosity}
\end{figure}
\begin{figure}
    \centering
    \includegraphics[scale=.45]{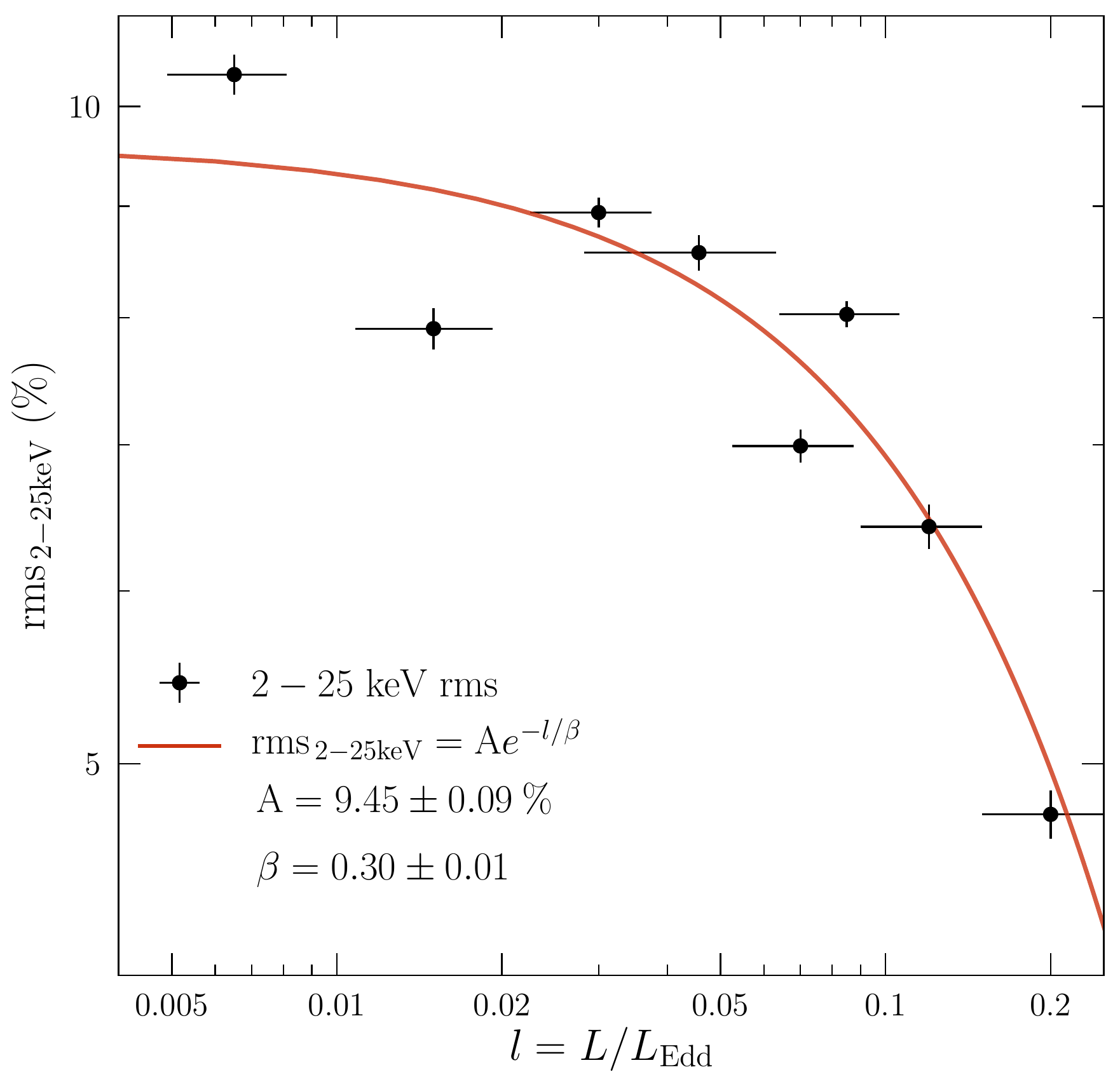}
    \caption{Total fractional rms amplitude, between 2 and 25 keV, of the lower kHz QPO vs. luminosity for the 8 sources in \cref{tab:lum_persource}. The red solid line indicates the best-fitting exponential model to the data (see text).}
    \label{fig:totalrms_luminosity}
\end{figure}

\section{Results}
\label{sec:results}
In \cref{tab:bestfit-lag/rms_vs_energy} we show the data we obtained from \citet{troyerSystematicSpectralTimingAnalysis2018}. The best fitting slope, $m$, and intercept, $c$, in the table are taken from \citet{troyerSystematicSpectralTimingAnalysis2018}, who fitted a linear relation to the energy-dependent time lags\footnote{Since most slopes in \citet{troyerSystematicSpectralTimingAnalysis2018} are negative, here we reversed the sign of the slope in the linear relation so that most of the slopes are positive.} in Fig. 8 and 9, $\Delta t(E)$ vs. energy, $\Delta t(E) = -mE + c$. The sources in \cref{tab:bestfit-lag/rms_vs_energy} are sorted on the basis of this best-fitting slope $m$. We also show in \cref{tab:bestfit-lag/rms_vs_energy} the total fractional rms amplitude integrated from 2 to 25 keV for both the lower and upper kHz QPOs and the best-fitting parameters of a broken-line model fit we performed to the fractional rms amplitude of the lower kHz QPO as a function of energy \citep[see Fig. 4 in][]{troyerSystematicSpectralTimingAnalysis2018}, with slopes $\alpha_1$ and $\alpha_2$, below and above the break energy, $E_{\mathrm{break}}$, respectively. Because for some sources $E_{\mathrm{break}}$ and $\alpha_1$ and/or $\alpha_2$ were not well constrained (e.g. 4U 1915$-$05, SAX J1748.9$-$202 and 4U 1705$-$44), while in all cases the slope above the break was consistent with zero within errors, in the fit we fixed $\alpha_2$ to zero and linked $E_{\mathrm{break}}$ to be the same for all sources.

From \cref{tab:bestfit-lag/rms_vs_energy} it is apparent that the slope below the break energy of the fractional rms amplitude fit generally decreases as the slope of the lag spectrum decreases. If we consider only the sub-sample of 8 sources with data over the full QPO frequency range, for which we have obtained luminosities from the literature (listed in \cref{tab:lum_persource}), it is apparent that as the luminosity increases the slope of the fractional rms spectrum and that of the time-lag spectrum decrease, indicating that both quantities depend upon luminosity.

To explore this, in \cref{fig:slopelag_luminosity} we plot the slope of the time lag spectrum (from \cref{tab:bestfit-lag/rms_vs_energy}) of the lower kHz QPO vs. luminosity for the 8 sources in our sub-sample (see \cref{tab:lum_persource}). The solid red line in the figure corresponds to the best-fitting exponential model, $m = Ae^{-l/\beta}$ where $l = L/L_\mathrm{Edd}$, to the data, with $A = 7.6 \pm 0.6$ $\mu$s/keV and $\beta = 0.08 \pm 0.01$. In this figure it is apparent that as the luminosity increases the slope in the time-lag spectrum decreases exponentially. The same behaviour can be observed in \cref{fig:totalrms_luminosity}, were we plot the total fractional rms amplitude integrated from 2 to 25 keV of the lower kHz QPO vs. luminosity for the same sub-sample of sources. Like in \cref{fig:slopelag_luminosity}, the solid red line corresponds to the best-fitting exponential model to the data, rms$_{2-25\mathrm{keV}} = Ae^{-l/\beta}$, with values $A = 9.45 \pm 0.09$ \% and $\beta = 0.30 \pm 0.01$. As in the case of the slope of the time-lag spectrum, the total 2-25 keV rms amplitude also decreases exponentially with the luminosity of the source. 

Although we do not show it here, the average lag between two broad energy bands \citep[see values in][Table 2 and 3]{troyerSystematicSpectralTimingAnalysis2018} and the slope below the break of the rms spectrum (see values in \cref{tab:bestfit-lag/rms_vs_energy}) of the lower kHz QPO behave in the same way when plotted vs. the luminosity of the source, with those quantities also decreasing exponentially with increasing luminosity.

\begin{figure}
    \centering
    \includegraphics[scale=.45]{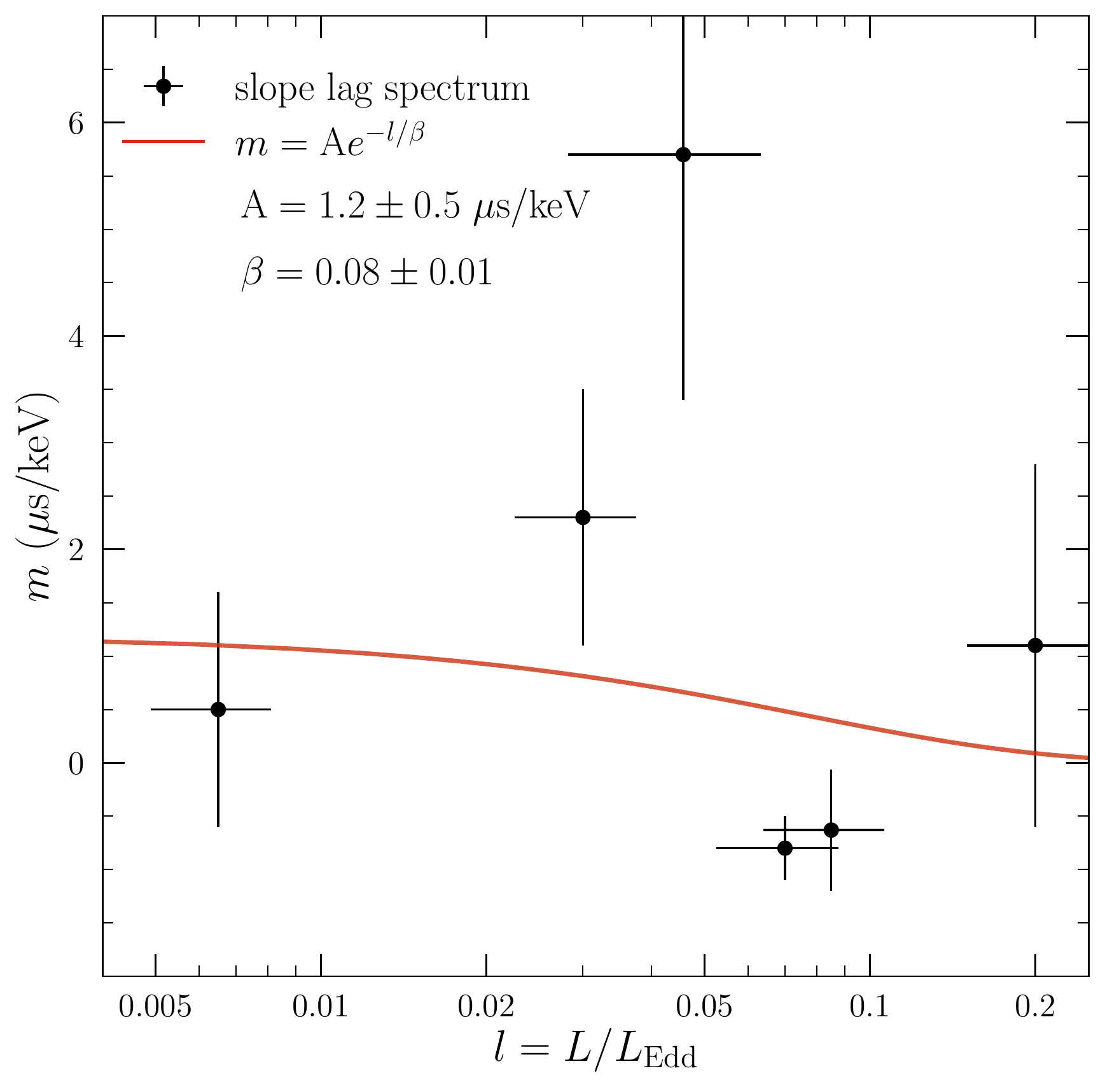}
    \caption{Slope of the time-lag spectrum, $m$ \citep[taken from][]{troyerSystematicSpectralTimingAnalysis2018}, of the upper kHz QPO vs. luminosity in Eddington units for the 6 sources in \cref{tab:lum_persource} that show upper kHz QPOs. The red solid line indicates the best-fitting exponential model to the data (see text).}
    \label{fig:slopelag_luminosity_upper}
\end{figure}

\begin{figure}
    \centering
    \includegraphics[scale=.45]{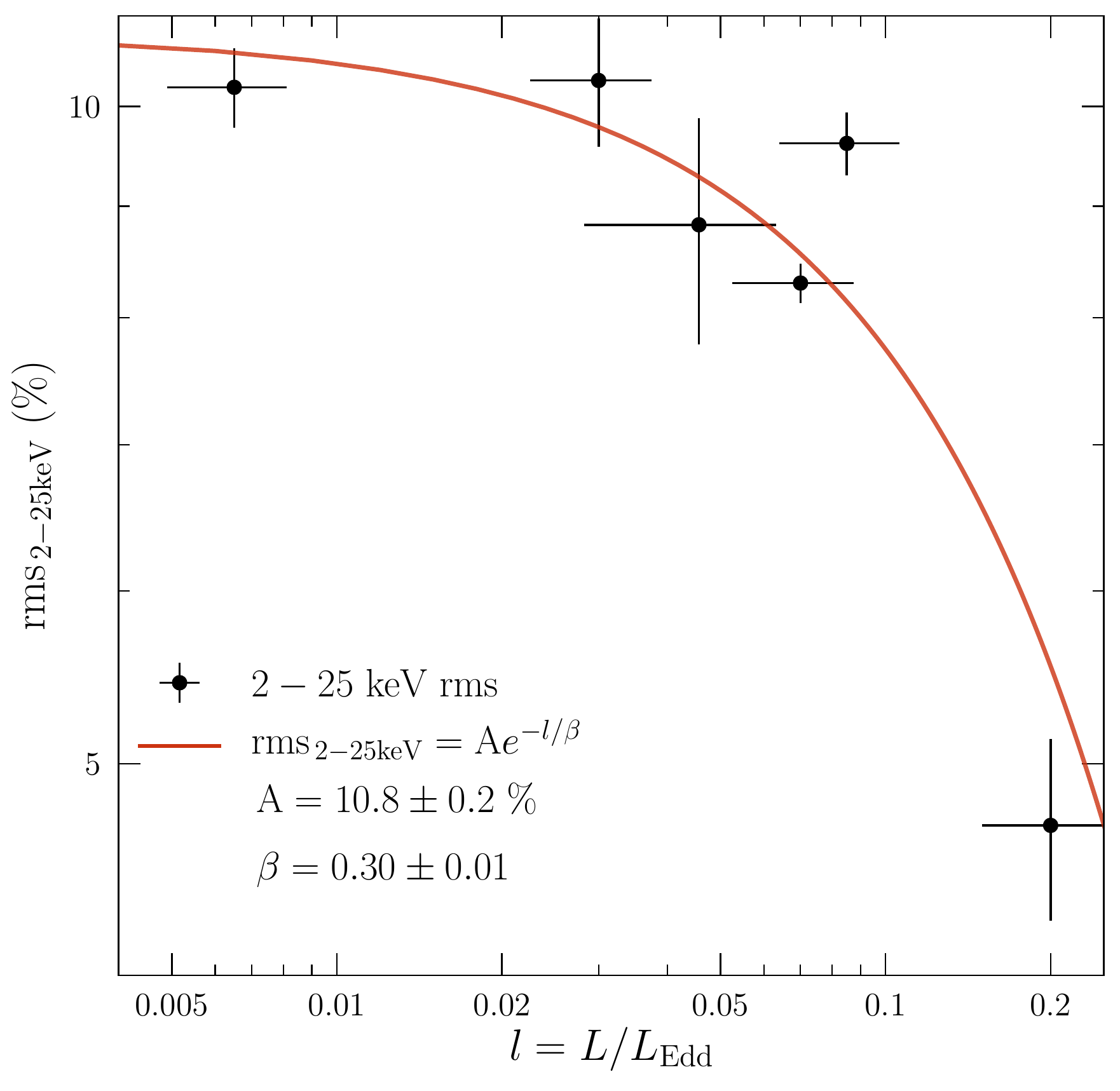}
    \caption{Total fractional rms amplitude, between 2 and 25 keV, of the upper kHz QPO vs. luminosity for the 6 sources in \cref{tab:lum_persource} that show upper kHz QPOs. The red solid line indicates the best-fitting exponential model to the data (see text).}
    \label{fig:totalrms_luminosity_upper}
\end{figure}

We performed a similar analysis for the relations of the slope of the time-lag spectrum and the total fractional rms amplitude upon luminosity on the 6 sources of \citet{troyerSystematicSpectralTimingAnalysis2018} where they observed the presence of the upper kHz QPOs (see \cref{tab:bestfit-lag/rms_vs_energy}). In \cref{fig:slopelag_luminosity_upper} we show the slope of the time-lag spectrum of the upper kHz QPO vs. luminosity, with the red solid line showing the best-fitting exponential model, $m = Ae^{-l/\beta}$, to the data. Due to the time-lags being less well constrained for the upper than the lower kHz QPO, we performed a joint fit of the exponential model on the time-lag slope data of both the lower and the upper kHz QPOs. In this joint fit we linked $\beta$ to be the same for both data-sets, as these parameters were consistent with each other in an independent fit. The resulting best-fitting parameters for the joint model are $A = 1.2 \pm 0.5$ $\mu$s/keV and $\beta = 0.08 \pm 0.01$. Because the value of $A$ in this model is consistent with zero at 2.2$\sigma$, we performed an F-test to determine whether a model with $A$ free would be favoured over a model with $A$ fixed to zero. The F-test probability is 0.5, which means that the slope of the time-lag spectrum of the upper kHz QPO is consistent with zero. In \cref{fig:totalrms_luminosity_upper} we show the total fractional rms amplitude of the upper kHz QPO vs. luminosity. Here we fit again an exponential model, rms$_{2-25\mathrm{keV}} = Ae^{-l/\beta}$, jointly to both the lower and upper kHz QPOs data-sets with $\beta$ linked. The best-fitting parameters resulting from the joint fit are $A = 10.8 \pm 0.02$ \% and $\beta = 0.30 \pm 0.01$. As is the case for the lower kHz QPO, here the total fractional rms amplitude also decreases exponentially with increasing luminosity.

Since both the fractional rms amplitude and the slope of the time-lag energy spectrum of the lower kHz QPO correlate with luminosity in a similar fashion, in \cref{fig:slopelag_totalrms} we plot these quantities together for the same sub-sample of sources in \cref{fig:slopelag_luminosity} and \cref{fig:totalrms_luminosity}. The red solid line in the figure corresponds to the best-fitting linear model, $m = a*\mathrm{rms}_{2-25\mathrm{keV}} + b$, with $a = 1.1 \pm 0.3$ $\mu$s/(keV\%) and $b = -4.8 \pm 2.1$ $\mu$s/keV. From this figure it is apparent that there is a strong linear correlation between the slope of the time-lag spectrum and the total rms amplitude of the sources. Not shown here, there are similar strong correlations between the broadband lag and both the total 2-25 keV rms amplitude and the slope below the break of the rms spectrum.

\section{Discussion}
\label{sec:discussion}
We discovered that, as with the rms fractional amplitude, the energy-dependent time lags of the lower kHz QPOs depend upon luminosity in 8 neutron-star LMXBs. Indeed we showed, for the first time, that the slope of the time-lag spectrum of the lower kHz QPO decreases exponentially with increasing luminosity. Since the fractional rms amplitude of the lower kHz QPO also decreases exponentially with luminosity, it follows that the slope of the time-lag spectra increases linearly with the average rms amplitude of the lower kHz QPO. For the upper kHz QPOs detected in 6 of the the 8 neutron-star LMXBs we studied, the fractional rms amplitude decreases exponentially with luminosity, while the slope of the time-lag spectrum is consistent with zero.

\begin{figure}
    \centering
    \includegraphics[scale=.45]{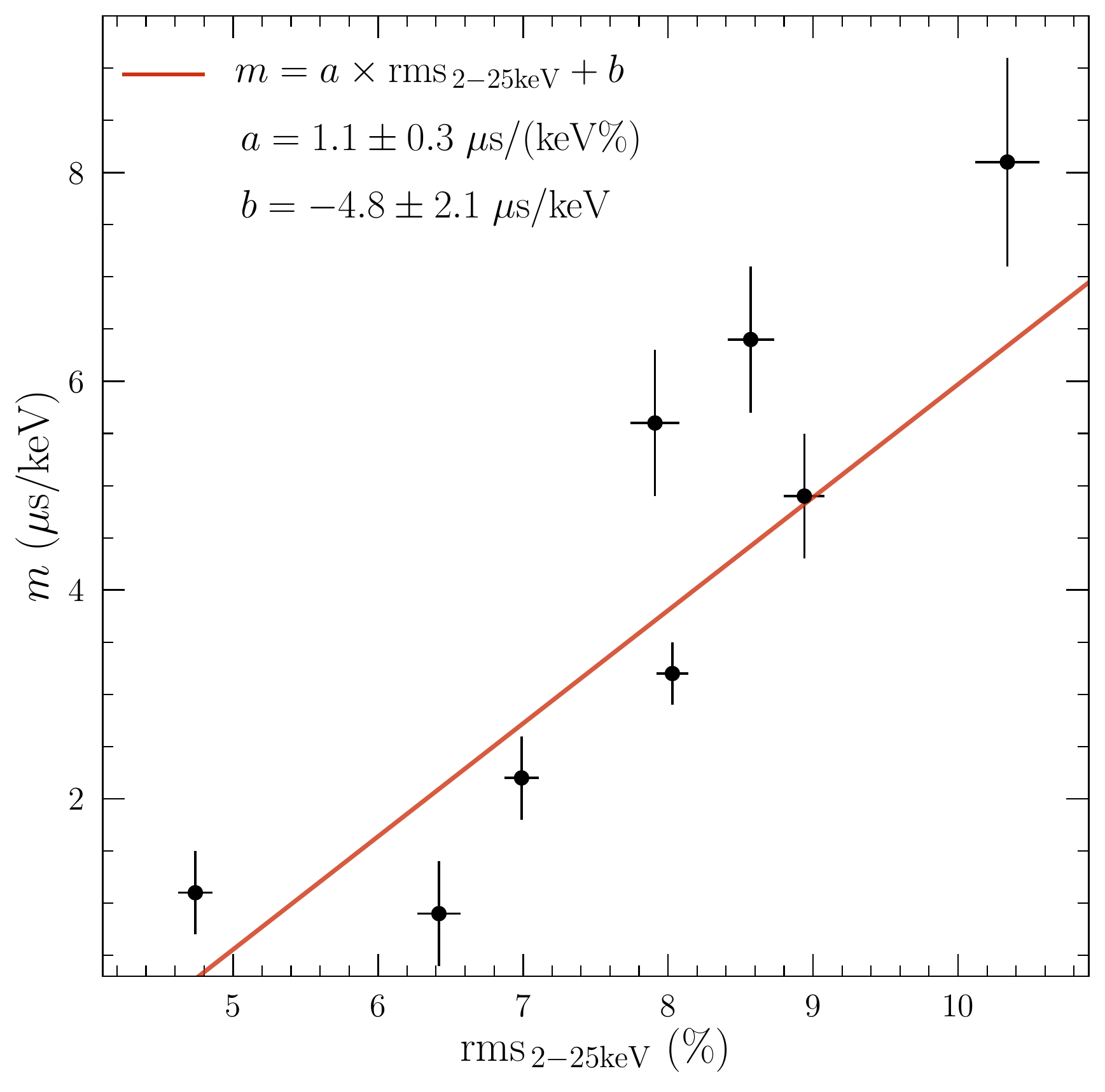}
    \caption{Slope of the time lag, $m$, versus the total fractional rms amplitude, between 2 and 25 keV, of the lower kHz QPO for the 8 sources in \cref{tab:lum_persource}. The red solid line indicates the best-fitting linear model to the data (see text).}
    \label{fig:slopelag_totalrms}
\end{figure}

The lags of the lower kHz QPOs are soft, meaning that the variability in the soft band is lagging behind that in the hard band. High energy X-ray photons reflected off an accretion disc have been observed to produce soft lags, known as reverberation lags, in black-hole LMXBs \citep[see][for a review of reverberation in accreting black holes]{uttleyXRayReverberationAccreting2014}. Particularly for neutron-star systems, \citet{cackettAreKHzQPO2016} studied the soft lags of the lower kHz QPO in 4U 1608$-$52 and explored the possibility that the lags were due to reverberation \citep[][]{barretSOFTLAGSNEUTRON2013}. \citet{cackettAreKHzQPO2016} showed that the lag spectrum should increase with energy above 8 keV if reverberation is the sole mechanism responsible for the soft lags of the lower kHz QPO in 4U 1608$-$52. Observations show the opposite trend, suggesting that reverberation is not the mechanism producing the soft lags of the lower kHz QPOs in 4U 1608$-$52; by extension, the same must be true for the lower kHz QPOs in the other sources in this paper. It is worth mentioning that \citet{coughenourModelingUpperKHz2020} performed a similar study of the lags of the upper kHz QPO in the source 4U 1728$-$34, finding that reverberation alone cannot be responsible for those lags either.

To make the case stronger against the accretion disc being directly responsible for the soft lags we observe in these sources, we can consider the behaviour of the fractional rms amplitude for the 14 sources in Fig. 4 in \citet{troyerSystematicSpectralTimingAnalysis2018}. The fractional rms amplitude of the kHz QPOs in these 14 sources increases with energy up to 10 keV, reaching values of 10-12\% at $\sim20$ keV \citep[see also][where similar results are shown]{bergerDiscovery800HZ1996,gilfanovBoundaryLayerAccretion2003,ribeiroAmplitudeKilohertzQuasiperiodic2019}. At these energies, the spectrum of a neutron-star LMXB is dominated by the emission of a Comptonising component or corona \citep[][]{sunyaevComptonizationXraysPlasma1980,whiteXraySpectralProperties1988}, while the contribution of the accretion disc is negligible, suggesting that the corona must be responsible for the amplitude and the lag spectra of the lower kHz QPOs \citep[][]{sannaKilohertzQuasiperiodicOscillations2010}. 

Inverse Compton scattering in the corona, as the electrons transfer energy to the soft photons coming from the accretion disc or the surface of the neutron star, produces a time delay of the hard photons with respect to the directly emitted soft photons \citep[][]{wijersEnergyDependentDelay1987,leeComptonizationQPOOrigins1998}. Inverse Compton scattering would therefore produce only hard lags, in conflict with the observations of soft lags in our data. However, if we consider Comptonisation as proposed by \citet{leeComptonUpscatteringModel2001}, \citet{kumarConstrainingSizeComptonizing2016} and \citet{karpouzasComptonizingMediumNeutron2020}, we can explain the observed soft lags accounting for feedback occurring between the corona and the accretion disc. In this scenario a fraction of the Comptonised photons impinge back onto the accretion disc, where they are thermalised and emitted at later times and lower energies than the photons coming from the corona. Since this effect produces another time delay, now for the soft photons that returned to the disc, we would observe soft lags, which is what we find in the data \citep[see][]{kumarEnergyDependentTime2014,karpouzasComptonizingMediumNeutron2020}.

In \cref{fig:slopelag_totalrms} both the total rms amplitude and the slope of the time-lag energy spectrum of the lower kHz QPO are shown to be linearly correlated with each other, as they both decrease in a similar manner with increasing luminosity. This trend suggests that there is a single property of the system that drives the changes we observe in the rms amplitude and the time-lags of the kHz QPOs. The dependence upon luminosity could imply that the trends we see in \cref{fig:slopelag_luminosity}, \cref{fig:totalrms_luminosity} and \cref{fig:totalrms_luminosity_upper} are determined by changes in the corona, as the sources become softer due to a smaller coronal contribution to the total energy spectrum with increasing luminosity. At low luminosities, for example, the accretion rate decreases and the energy spectrum becomes hard, the corona is then optically thin and the amplitude of the variability increases. In this scenario the properties of the corona drive the properties of the variability, as it occurs in the Compton up-scattering model we described in the previous paragraph.

As we explained in \cref{sec:data}, the luminosity of the source can change significantly (by a factor of $\sim1.5-5$ depending on the source) when the lower kHz QPO is present; the spread, however, is $\sim10-30$\% of the average luminosity. On the other hand, when the lower kHz QPO is present the hard colour of the source changes by less than $\sim5-10$\% peak to peak \citep[see e.g., Fig. 2 and 6][]{barretSupportingEvidenceSignature2007}, whereas at the vertex of the colour-colour diagram, when the lower kHz QPO is present, the average hard colour of these sources changes by a factor $\sim2$ \citep[see e.g.,][]{vanstraatenRelationsTimingFeatures2000,zhangCoolingPhaseType2011,garciaCoolingPhaseType2013,zhangRelationSpectralChanges2017}. If the hard colour is a proxy of the properties of the corona, the above considerations show that the change of the average properties of the corona from source to source is significantly larger than the change within a single source when the lower kHz QPO is present. Since the hard colour and luminosity of neutron-star X-ray binaries are correlated \citep[][]{vanparadijsLuminosityDependenceHardness1994}, it is appropriate to take the luminosity as a proxy for the properties of the corona.

To further understand the dependence of the properties of the kHz QPOs upon the properties of the corona, we can examine what happens with the quality factor as a function of the frequency of the QPO and the luminosity of the source. Similar to the fractional rms amplitude, the quality factor of the lower kHz QPO increases and then decreases with QPO frequency \citep[see e.g.,][]{disalvoCorrelatedSpectralTiming2003,barretDropCoherenceLower2005} and with increasing luminosity \citep[see e.g.,][]{mendezMaximumAmplitudeCoherence2006}. The fact that the quality factor, the fractional rms amplitude and the time lags of the lower kHz QPO depend upon luminosity suggests that there is a coupled mode of oscillation between the corona and the disc. \citet{leeComptonUpscatteringModel2001} proposed that this mode is responsible for the lower kHz QPO, which would arise from a resonance between oscillations in the disc (or the neutron star surface) and the corona \citep[see e.g.,][]{deavellarTimeLagsKilohertz2013,deavellarPhaseLagsQuasiperiodic2016,zhangRelationSpectralChanges2017,ribeiroRelationPropertiesKilohertz2017,ribeiroAmplitudeKilohertzQuasiperiodic2019}. The frequency at which the quality factor and the rms amplitude of the QPO peak, would correspond to a natural frequency of the corona that depends upon its properties. At high luminosities, the contribution of the corona to the energy spectrum of a source and the feedback to the disc decreases and the oscillation of both the disc and the corona is not in resonance, leading to a decrease of the rms amplitude, quality factor and lags.

The nature of the oscillation mode we believe to be responsible of the variability we observe in the emission of the neutron-star LMXBs in our sample, and the behaviour of the timing properties of the QPOs, is uncertain. To better comprehend what drives this oscillation resonance, we can consider the work of \citet{karpouzasComptonizingMediumNeutron2020}, based on the model in \citet{kumarEnergyDependentTime2014}, where they conceive the QPO as a small oscillation in the solution of the Kompaneets equation, that describes the evolution of the energy spectrum of the source dominated by inverse Compton scattering \citep[][]{kompaneetsEstablishmentThermalEquilibrium1957}. The model in \citet{karpouzasComptonizingMediumNeutron2020} links the behaviour of the timing properties of the QPO to the physical quantities accounted for in the Kompaneets equation, e.g. the electron temperature and size of the corona. \citet{karpouzasComptonizingMediumNeutron2020} fit their model to both the rms and lag spectra of the lower kHz QPO in 4U 1636$-$53 and find that the perturbations in coronal properties are the ones producing the oscillations in the emission of the source, reaching their maximum variability at a QPO frequency of 700 Hz. These results suggest that the corona is responsible for the dependence upon QPO frequency of the lag and the fractional rms amplitude of the lower kHz QPO, and hints at the existence of a resonance between the source of the soft photons and the corona. This conclusion agrees with what we find in the present paper: the luminosity of the source is related to properties of the corona, like its temperature, optical depth and size, that together with the feedback fraction of photons that return to the soft-photon source, can explain the energy-dependent fractional rms amplitude and time-lags of the lower kHz QPOs \citep[see Section 2 in][for a more detailed explanation of the model]{karpouzasComptonizingMediumNeutron2020}.

Compton up-scattering model of \citet{leeComptonUpscatteringModel2001} could also explain the difference between the properties of the time-lag of the lower and upper kHz QPOs (see \cref{fig:slopelag_luminosity}, \cref{fig:slopelag_luminosity_upper} and \cref{tab:bestfit-lag/rms_vs_energy}), due to the dependence upon frequency of the resonance between the disc and the corona. In neutron-star LMXBs the coexistence between the soft lags of the lower kHz QPOs and the hard lags of the upper kHz QPOs \citep[see e.g.,][]{deavellarTimeLagsKilohertz2013,peilleSpectraltimingPropertiesUpper2015} can be expected, because the oscillations of the coronal properties depend on QPO frequency in this Compton up-scattering model. Similarly, the presence of other features, like the hard lags of the BBN component that remain roughly independent of frequency in the range between $\sim$0.01 to $\sim$100 Hz \citep[see e.g.,][]{miyamotoDelayedHardXrays1988,fordMeasurementHardLags1999}, could be also predicted by the model.

In the context of the scenario that we described in this section, the neutron-star LMXB XTE J1701$-$462 becomes a unique case to study, as is one of only two sources to show atoll-like and Z-like\footnote{Neutron-star LMXBs are usually divided into two categories: atoll and Z, depending on the pattern they describe during an outburst in their colour-colour diagram \citep{hasingerTwoPatternsCorrelated1989}.} behaviour during the same outburst \citep[][]{homanRossiXRayTiming2007,linSpectralStatesXTE2009,homanXTEJ17014622010}. \citet{sannaKilohertzQuasiperiodicOscillations2010} studied the kHz QPOs in XTE J1701$-$462, both in the atoll and Z phases, and found that the maximum quality factor and the maximum rms amplitude of the lower kHz QPOs of both phases follow the same trend with luminosity that \citet{mendezMaximumAmplitudeCoherence2006} described for other neutron-star LMXBs \citep[see Fig. 5 in][]{sannaKilohertzQuasiperiodicOscillations2010}. The results presented here suggest that the time-lag energy spectrum of the lower kHz QPO in the two different phases of XTE J1701$-$462 will change with luminosity in a similar way to what we show in \cref{fig:slopelag_luminosity}. Such a relation, if observed, would help to further understand the nature of the mechanism responsible for the lower kHz QPO in neutron-star LMXBs, isolating the changes in the system that luminosity depends upon from other factors like the mass of the neutron star, the inclination of the accretion disc or the strength of the magnetic field.

\section*{Acknowledgements}

The authors wish to thank Federico Garc\'ia and Konstantinos Karpouzas for useful discussions that helped with the ideas presented in this manuscript. We also thank the referee for insightful comments that helped improve the clarity of the paper. This research has made use of data obtained from the High Energy Astrophysics Science Archive Research Center (HEASARC), provided by NASA’s Goddard Space Flight Center.

\section*{Data Availability}
The data underlying this article are publicly available at the website of the High Energy Astrophysics Science Archive Research Center (HEASARC, \url{https://heasarc.gsfc.nasa.gov/}).



\bibliographystyle{mnras}
\bibliography{main}




\bsp	
\label{lastpage}
\end{document}